\documentclass{Interspeech2024}
\usepackage{verbatim}

\interspeechcameraready

\title{Inclusive ASR for Disfluent Speech: Cascaded Large-Scale Self-Supervised Learning with Targeted Fine-Tuning and Data Augmentation}

\name[affiliation={1}]{Dena}{Mujtaba}
\name[affiliation={1}]{Nihar R.}{Mahapatra}
\name[affiliation={1}]{Megan}{Arney}
\name[affiliation={1}]{J. Scott}{Yaruss}
\name[affiliation={2}]{Caryn}{Herring}
\name[affiliation={1}]{Jia}{Bin}
\address{
  $^1$Michigan State University\\
  $^2$FRIENDS: The National Association of Young People Who Stutter}
\email{\{mujtabad,nrm,arneymeg,jsy,binjia\}@msu.edu, caryn@friendswhostutter.org}


\begin{document}

\maketitle

\begin{abstract}
Automatic speech recognition (ASR) systems often falter while processing stuttering-related disfluencies---such as involuntary
blocks and word repetitions---yielding inaccurate transcripts.  A critical barrier to progress is the scarcity of large, annotated disfluent speech datasets. Therefore, we present an inclusive ASR design approach, leveraging large-scale self-supervised learning on standard speech followed by targeted fine-tuning and data augmentation on a smaller, curated dataset of disfluent speech. Our data augmentation technique enriches training datasets with various disfluencies, enhancing ASR processing of these speech patterns. Results show that fine-tuning wav2vec 2.0 with even a relatively small, labeled dataset, alongside data augmentation, can significantly reduce word error rates for disfluent speech. Our approach not only advances ASR inclusivity for people who stutter, but also paves the way for ASRs that can accommodate wider speech variations.
\end{abstract}

\section{Introduction}
\textit{Stuttering} is a complex neurodevelopmental condition characterized by moments of involuntary disruptions in the flow of speech when people who stutter experience a loss of control \cite{tichenor2019stuttering}. These disruptions manifest as \textit{disfluencies} like word repetitions (e.g., ``my my my name is''), prolongations (e.g., ``mmmy name is''), and blocks or pauses (e.g., ``my n---ame is'') \cite{wu2023world,johnson1959onset}. Although these and other types of disfluencies occur in everyday speech, they are markedly more frequent in individuals who stutter, a condition that affects over 80 million people globally \cite{wu2023world,yairi2013epidemiology}. The prevalence and severity of these disfluencies can vary significantly across individuals, influenced by environmental and situational factors \cite{tichenor2021variability}.

The widespread integration of automatic speech recognition (ASR) into voice-activated artificial intelligence (voice AI) applications across sectors such as education, employment, home automation, and transportation presents a significant challenge for people who stutter. Although ASRs can achieve up to 95\% word accuracy on standard speech \cite{tobin2022personalized}, their performance markedly declines when processing disfluent speech, often failing to accurately decode the variations inherent in stuttering \cite{wu2023world,lea2023user}. This discrepancy not only impacts people who stutter in their daily and professional lives but also amplifies societal biases. It leads to discrimination, restricts equal opportunities in automated services and employment, and contributes to difficulties in utilizing voice-activated services like automated phone systems, voice assistants (e.g., Apple's Siri, Amazon's Alexa), and biases in automated job interview scoring systems \cite{mujtaba2019ethical}. The effect of these biases is profound, often resulting in decreased job satisfaction and increased marginalization, as fluent speech is frequently preferred in employment settings \cite{gerlach2018stuttering,plexico2019influence}. Consequently, the disparity in ASR accuracy will exacerbate the marginalization of people who stutter, impacting all downstream applications.

\subsection{Related Work}

Addressing the need for inclusive ASR models has prompted diverse strategies. One such approach involves providing alternative communication options, like typing commands for voice assistants \cite{techexplore}. While helpful, this method does not directly tackle the intrinsic bias within ASRs. In a more direct effort, initiatives such as Google's Project Euphonia and Relate have sought to refine the core ASR model, collecting diverse speech data to better accommodate conditions like dysarthria \cite{tobin2022personalized}. Nonetheless, these efforts have not specifically addressed the challenges presented by stuttering-related disfluent speech, characterized notably by its variability \cite{tichenor2021variability}.

In response to the distinctive needs of people who stutter, research has increasingly focused on creating ASR systems specifically designed for stuttering. The majority of such work has focused on disfluency detection, wherein disfluency events are detected and classified and subsequently adjusted, ignored, highlighted, or removed from speech prior to ASR processing, thereby enhancing accuracy \cite{sheikh2022machine}. Although datasets like SEP-28k and LibriStutter have emerged to support disfluency detection \cite{lea2021sep, kourkounakis2020fluentnet}, there is a scarcity of datasets with paired transcripts and audio to train ASR models. Very few approaches have been proposed to tailor ASR models directly for stuttering. Lea et al., for instance, tuned Apple Speech framework model parameters using data on stuttering characteristics (e.g., distribution of duration of blocks) obtained from individuals who stutter \cite{lea2023user}. Similarly, Mitra et al. focused on tuning decoding parameters of ASRs \cite{mitra2021analysis}. Other approaches have also been investigated to enhance ASR models, although they focus on other types of speech variations like dysarthria, dysphonia, or aphasia \cite{tobin2022personalized,moore2020uncommonvoice,hermannfew,baskar2022speaker,macwhinney2024talkbank}.

\subsection{Our Contributions}
This paper aims to address the aforementioned challenges and develop an accessible ASR for people who stutter using a combination of fine-tuning and data augmentation. A primary limitation in creating an ASR for stuttered speech is the absence of a comprehensive dataset covering various types of disfluencies and their variations among individuals who stutter. To overcome this limitation, we introduce a novel speech data augmentation method specifically designed for stuttered speech that also allows for the manipulation and control of disfluency types during the fine-tuning process.

The core contributions of this paper, distinguishing it from prior efforts, are fourfold: (1) ASR fine-tuning for accessibility with a stuttering-specific focus: We investigate the impact of fine-tuning a pre-trained wav2vec 2.0 model with stuttered speech datasets of various sizes on ASR word error rate (WER) and semantic similarity, aiming to enhance model performance for disfluent speech. We specifically target disfluencies in stuttering, which vary greatly across people and environments, necessitating specialized solutions. (2) Disfluent speech data augmentation: An innovative data augmentation method designed specifically for stuttered speech is introduced, overcoming the challenges of limited data. This method provides precise control over the types, frequency, and placement of disfluencies within speech samples, enabling the enrichment of our dataset and more effective training of robust ASRs. (3) Accuracy bias analysis: We evaluate the effectiveness of our method in mitigating accuracy bias---the discrepancy in ASR performance when processing speech with and without stuttering. By comparing ASR performance against a baseline of non-stuttered speech, we assess the effectiveness of our strategies in promoting equitable ASR accuracy. (4) Diverse and realistic evaluation settings: We evaluate our fine-tuned ASR using speech from diverse contexts, including interview and reading videos, and assess its performance across various demographics of people who stutter. With these contributions, we anticipate our approach will advance the development of fair and accessible ASRs for individuals who stutter, with potential extensions to accommodate other speech differences featuring disfluencies.

\section{Methodology}
\begin{figure}[t!]
\centering
\includegraphics[width=.99\columnwidth]{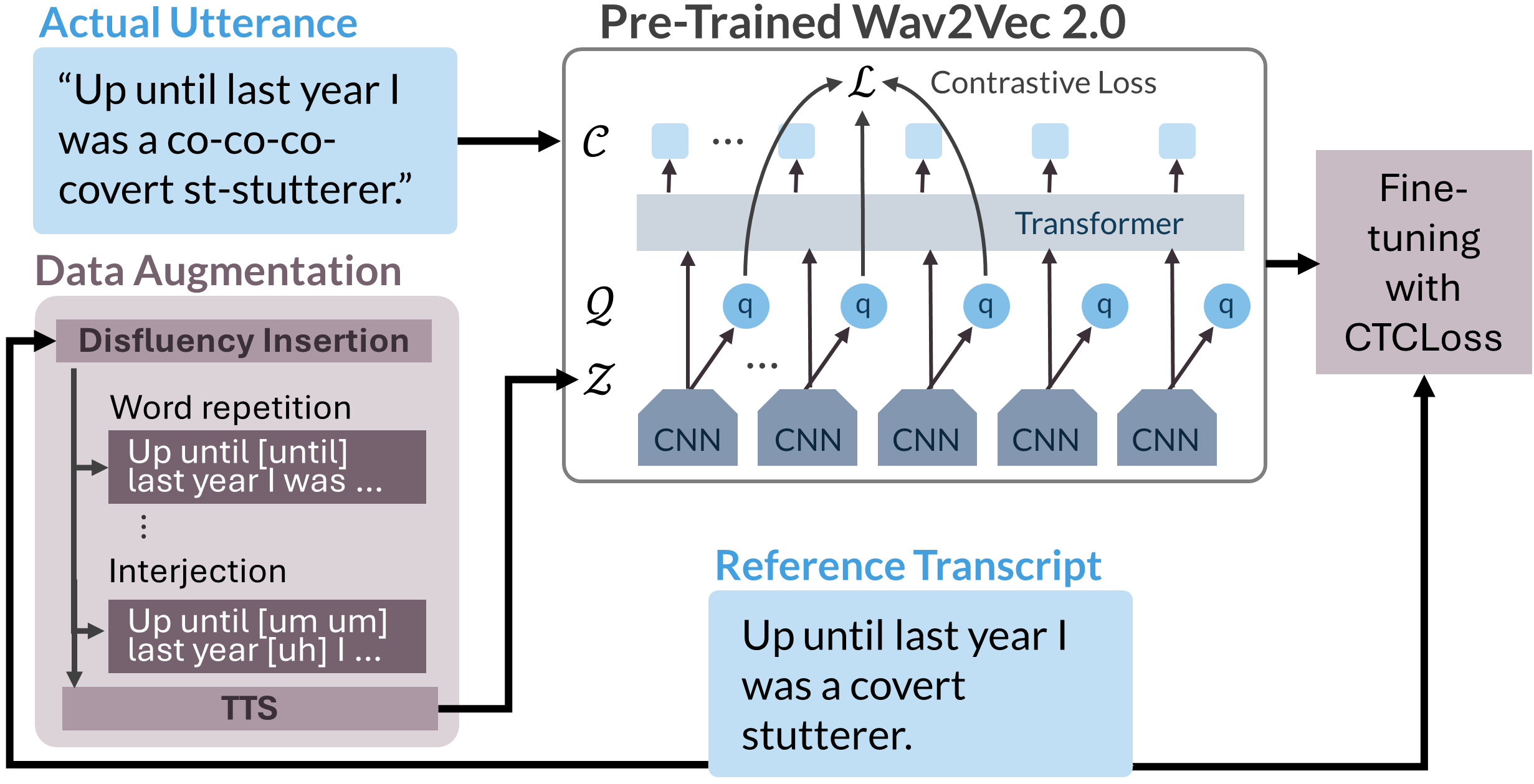}
\caption{Schematic of our method integrating data augmentation and fine-tuning of wav2vec 2.0 for stuttered speech, exemplified by augmentations from a FluencyBank speech sample \cite{ratner2018fluency}. In wav2vec 2.0, $\mathcal{L}$ denotes the loss function, $\mathcal{C}$ represents context representations, $\mathcal{Q}$ indicates quantized representations, and $\mathcal{Z}$ corresponds to latent speech representations.}
\label{fig:overview}
\vspace*{-.2in}
\end{figure}

\subsection{wav2vec 2.0}
Our study leverages wav2vec 2.0, a large-scale, self-supervised model designed for speech processing. Figure \ref{fig:overview} provides a visual overview of the model's architecture. Pre-trained on an extensive corpus of unlabeled speech, it can be subsequently fine-tuned on a smaller, targeted dataset. This model excels in constructing a contextually rich representation of speech through a series of lightweight components. Compared to larger end-to-end models such as Whisper \cite{radford2023robust}, wav2vec 2.0, although not a state-of-the-art model, offers a more computationally-efficient approach for our experiments, providing a balance of performance and feasibility. Additionally, it offers precise control over the training dataset, avoiding the potential for data leakage into test sets that could occur with Whisper due to its opaque training data. For fine-tuning, we utilize the `wav2vec2-base-960h' model variant from HuggingFace \cite{wav2vecweights}, pre-trained on 960 hours of the LibriSpeech dataset \cite{panayotov2015librispeech}. This selection aids in precisely attributing performance enhancements to our methodological innovations, despite the model's potential suboptimal performance on conversational speech.

\subsection{FluencyBank}
For assessing ASR performance on authentic stuttered speech, we utilize the FluencyBank dataset \cite{ratner2018fluency}, which comprises video recordings of people who stutter in two distinct settings: reading passages and participating in interviews. This collection features 7 readings and 12 interview videos from 12 participants, each paired with a transcript adhering to the Codes for the Human Analysis of Transcripts (CHAT) standard \cite{macwhinney2017tools}. CHAT transcripts segment speech into individual \textit{utterances}---sentences or fragments thereof, uttered by a single speaker---annotated with disfluency event labels. Faced with temporal misalignment and inaccuracies in FluencyBank's CHAT transcripts---a problem also reported by others \cite{sheikh2022machine,lea2021sep}---we had this dataset re-annotated in CHAT format by trained speech-language pathologists with expertise in stuttering, while ensuring inter-annotator agreement. In total, the final dataset consists of 1373 utterances and 2.21 hours of audio duration.

\subsubsection{Non-Disfluent Speech Datasets}
To evaluate our fine-tuned ASR's efficacy on non-disfluent speech, we generated a modified version of the FluencyBank dataset, termed FluencyBank-N, from which all apparent disfluencies have been removed. This variant was synthesized using the SpeechT5 text-to-speech (TTS) model \cite{ao-etal-2022-speecht5} in conjunction with a selection of pre-trained speaker vectors \cite{cmuvectors}, aiming to produce speech that mirrors the original transcripts but without disfluencies. By carefully selecting and employing 10 distinct pre-trained speaker vectors, we ensured a varied vocal representation. This approach allows us to precisely assess how our ASR model performs on the FluencyBank dataset under the hypothetical scenario where the speakers exhibit no disfluencies. 
We note that although synthetic speech is not identical to real speech, the performance of wav2vec 2.0 on this synthetic dataset is comparable to, albeit slightly worse than, its performance on the test set. This similarity, detailed in our results (Sec. \ref{S-results}), indicates that the synthetic dataset is a suitable metric for studying accuracy bias.

\subsection{Data Augmentation for Disfluent Speech}
To overcome the diversity shortfall in small stuttered speech datasets, our data augmentation strategy introduces a broader spectrum of disfluency events, more closely mirroring natural speech patterns observed in individuals who stutter. While the LibriStutter dataset \cite{kourkounakis2020fluentnet} represents a prior attempt at simulating stuttering through audio processing, it falls short in capturing the full complexity of the stuttering phenomena, notably in variability. In contrast, our method generates a richer assortment of disfluency events, including \textit{word repetitions} (e.g., ``my my my name is''), \textit{phrase repetitions} (e.g., ``my name my name my name is''), and \textit{interjections} (e.g., ``um um my name uh is''), inserted at randomized locations and frequencies during ASR training, thereby producing a simulation that more accurately reflects the inherent variability in disfluent speech of real individuals. Although our approach does not cover every type of stuttered disfluency, the selected event types are crucial. They enhance the realism of the training data and help study the effectiveness of our augmented dataset, ultimately improving ASR accuracy for all types of stuttered disfluencies.

During training, we augment FluencyBank utterances to assess the impact of incorporating $N$ augmented samples, employing random sampling from the original dataset with replacement. Starting with disfluency-free ground truth transcripts, we introduce specific disfluency events in text form. These texts are processed by the OpenAI TTS API \cite{openai} to synthesize speech with the inserted disfluencies, using voices randomly selected from the OpenAI library. This approach enables us to simulate a wide range of voices and disfluency events, along with their occurrence frequency in each utterance, facilitating a comprehensive evaluation of ASR performance.

We randomize disfluency frequencies and placements within set ranges: (1) for word repetition, we select between 1 to 3 words in each utterance to repeat 1 to 4 additional times; (2) for phrase repetition, we choose a phrase of 2 to 4 words and repeat it 1 to 3 times; (3) for interjections, we randomly insert either ``uh'' or ``um'' in 1 to 4 locations and repeat them 1 to 4 times. Initially, we explore the effectiveness of adding $N$ samples during training, for $N = 500, 1000, 2000, 3000$, represented as $p$ percent of the original dataset size. 
Subsequently, we change the number of repeats of disfluencies in our generation approach, and produce up to $N=6000$ augmented samples to assess the impact of increased disfluency variability on ASR performance. The new ranges used are the following: (1) for word repetition, each selected word is repeated 1 to 6 additional times; (2) for phrase repetition, each selected phrase is repeated 1 to 5 additional times; (3) for interjections, each interjection is repeated 1 to 7 times. The upper bound for all modification ranges is adjusted to fit the shortest length of the utterance (e.g., if there are only two words in an utterance, no more than two disfluency insertions can occur).

\subsection{Evaluation Metrics}
To assess the performance of our ASR models and examine accuracy bias, we employ two metrics: the \textit{word error rate (WER)} and \textit{BERTScore}. \textit{WER} is a standard metric in ASR evaluations, defined as:
$
{\rm WER} = \frac{S + D + I}{S + D + C},
$
where $S$, $D$, $I$, and $C$ denote the number of word substitutions, deletions, insertions, and correct words, respectively, compared to the reference transcript. A lower WER, approaching zero, is desirable, indicating higher word transcription accuracy.

\textit{BERTScore}, on the other hand, measures semantic similarity between the predicted and reference sentences, crucial for identifying significant changes in meaning caused by minor transcription errors. Given an ASR-produced transcript $x = \left \langle x_1, ..., x_k \right \rangle$ and a reference transcript $\hat{x} = \left \langle \hat{x}_1, ..., \hat{x}_k \right \rangle$, where $x_i$ and $\hat{x}_i$ are $i^{th}$ token embeddings (e.g., from BERT), BERTScore is calculated as follows \cite{tobin2022assessing,zhang2019bertscore}:
\begin{equation}
F_{BERT} = 2\frac{P_{BERT}\times R_{BERT}}{P_{BERT} + R_{BERT}},
\end{equation}
where $
R_{BERT} = \frac{1}{|x|}\sum_{x_i\in x}\max_{\hat{x}_j \in \hat{x}}(x_i^\textup{T} \cdot \hat{x}_j), $ and
$
P_{BERT} = \frac{1}{|\hat{x}|}\sum_{\hat{x}_i\in \hat{x}}\max_{x_j \in x}(x_i^\textup{T} \cdot \hat{x}_j)
$. Here, $F_{BERT}$ represents the F1 score, and we utilize the re-scaled version of BERTScore to obtain a range of $[-1, 1]$, where 1 would be semantically identical transcripts. We also utilize the ``microsoft/deberta-large-mnli'' model embeddings \cite{huggingface_deberta_large_mnli}, which offers more nuanced contextual representations than BERT.

\subsection{Implementation}
For our study, we fine-tuned wav2vec 2.0 models on the FluencyBank dataset, augmented with $N$ randomly selected samples to address the limited dataset size. We implemented a 6-fold cross-validation strategy, ensuring each fold contained utterances from two distinct speakers, with no speaker overlap across folds. This segmentation of FluencyBank into six parts based on speaker identity guarantees that any two folds are free from speaker overlap, whether across videos, utterances, or generated utterances. Such a methodological approach allows for a focused evaluation of ASR performance on voices not previously encountered during training, with augmentation applied only to utterances from speakers allocated to each specific fold.

The fine-tuning process leveraged the Connectionist Temporal Classification Loss (CTCLoss), a conventional choice for optimizing wav2vec 2.0 \cite{baevski2020wav2vec}. Training was conducted on an NVIDIA V100 GPU for about 55 epochs, or until no further reduction in loss was observed, employing early stopping to identify the most effective model configuration per fold. In addition, through experimentation with several hyperparameters, we set the learning rate to 6.25e-6, weight decay to 0.01, and use a batch size of 4 with gradient accumulation applied at each step. Additionally, to standardize input data, normalization procedures were applied to adjust casing and punctuation using the BasicTextNormalizer function from Whisper, available through the HuggingFace library \cite{normalizer}.

\section{Results \& Discussion}\label{S-results}

\begin{table}[t]
\begin{center}
  \caption{Comparative analysis of ASR performance metrics (WER and $F_{BERT}$) on FluencyBank (FB) and FluencyBank-N (FBN). ``Base'' denotes wav2vec 2.0 without any fine-tuning. Subsequent ASRs shown are trained with a $p$ percentage increase in the training dataset size through augmentation.}
\begin{tabular}{ l|l|l|l|l }
\toprule
& \multicolumn{2}{c|}{WER} & \multicolumn{2}{c}{$F_{BERT}$}   \\ \midrule
& \multicolumn{1}{c|}{FB} & \multicolumn{1}{c|}{FBN} & \multicolumn{1}{c|}{FB} & \multicolumn{1}{c}{FBN}  \\ \midrule
Base  & .4223& .1029               & .4700 & .8561 \\ 
$p=0$  & .2721 & \textbf{.0952}     & .6357 & \textbf{.8884} \\ 
$p=36$  & .2705 & .1306             & .6291 & .8063 \\ 
$p=73$  & .2668 & .1389             & .6332 & .7736 \\ 
$p=146$  & .2631 & .1436            & .6349 & .7590 \\ 
$p=218$  & .2624 & .1399            & .6367 & .7791 \\ 
$p=291$  & .2581 & .1318            & .6402 & .7885 \\ 
$p=364$  & .2569 & .1336            & .6390 & .7833\\ 
$p=437$  & \textbf{.2522} & .1248   & \textbf{.6566} & .8128 \\ 
\bottomrule
\end{tabular}
\label{tbl:metrics}
\vspace*{-.3in}
\end{center}
\end{table} 

\begin{table}[t]
\begin{center}
  \caption{Comparative analysis of ASR WER on FluencyBank (FB) and FluencyBank-N (FBN) per video setting (i.e., reading or interview). ``Base'' refers to wav2vec 2.0 without any fine-tuning. Subsequent ASRs shown are trained with a $p$ percentage increase in the training dataset size through augmentation.}
\begin{tabular}{ l|l|l|l|l }
\toprule
& \multicolumn{2}{c|}{WER Reading} & \multicolumn{2}{c}{WER Interview}   \\ \midrule
& \multicolumn{1}{c|}{FB} & \multicolumn{1}{c|}{FBN}  & \multicolumn{1}{c|}{FB} & \multicolumn{1}{c}{FBN} \\ \midrule
Base  & .3775 & .1045                   & .4279 & .1027 \\ 
$p=0$  & .1912 & .0795     & .2822 & \textbf{.0972}\\ 
$p=36$   & .1947 & .0938            & .2800 & .1352 \\ 
$p=73$   & .1888 & .0938            & .2766 & .1445 \\ 
$p=146$  & .1834 & .1027            & .2731 & .1487\\ 
$p=218$  & .1780 & .0800            & .2729 & .1473 \\ 
$p=291$  & .1673 & .0777            & .2694 & .1385 \\ 
$p=364$  & .1583 & .\textbf{0699}   & .2691 & .1415\\ 
$p=437$  & \textbf{.1625} & .0783   & \textbf{.2634} & .1306\\ 
\bottomrule
\end{tabular}
\label{tbl:videotypemetrics}
\vspace*{-.2in}
\end{center}
\end{table}

\begin{figure}[t]
\centering
\includegraphics[width=.95\columnwidth]{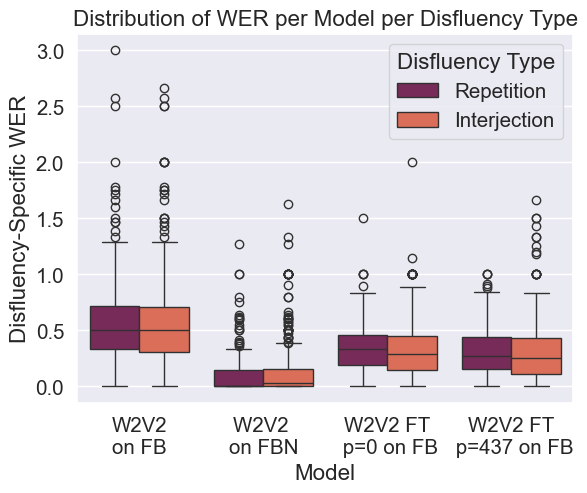}
\vspace*{-.1in}
\caption{Distribution of WER for four of our models per disfluency type: wav2vec 2.0 (W2V2) tested on FluencyBank (FB), W2V2 tested on FluencyBank-N (FBN), W2V2 fine-tuned (FT) on FB and tested on FB, and W2V2 fine-tuned with $N=6000$ additional samples (i.e., $p=437$) tested on FB.}\label{fig:dist_plot_disfluencies}
\vspace*{-.2in}
\end{figure}

We present the performance outcomes of the wav2vec 2.0 models on the FluencyBank dataset, both with disfluencies (FluencyBank) and without disfluencies (FluencyBank-N), in Table \ref{tbl:metrics}. The results reveal a consistent trend across all evaluated wav2vec 2.0 models: they achieve lower WER and higher $F_{BERT}$ scores  when tested on the FluencyBank-N dataset. This pattern suggests the presence of systemic bias in the models against stuttered speech.

Fine-tuning wav2vec 2.0 on datasets containing stuttering significantly improves both WER and $F_{BERT}$ metrics. Specifically, we observed a substantial 15\% reduction in WER for stuttered speech, even in the absence of augmented samples. Additionally, there is a 25\% enhancement in $F_{BERT}$, reflecting more accurate transcriptions that better capture the intended meaning. This fine-tuning not only benefits stuttered speech but also slightly enhances the performance on non-disfluent speech, as demonstrated by the FluencyBank-N dataset outcomes.

\begin{figure}[h]
\centering
\includegraphics[width=.8\columnwidth,trim={0cm 0cm 0cm 0cm},clip]{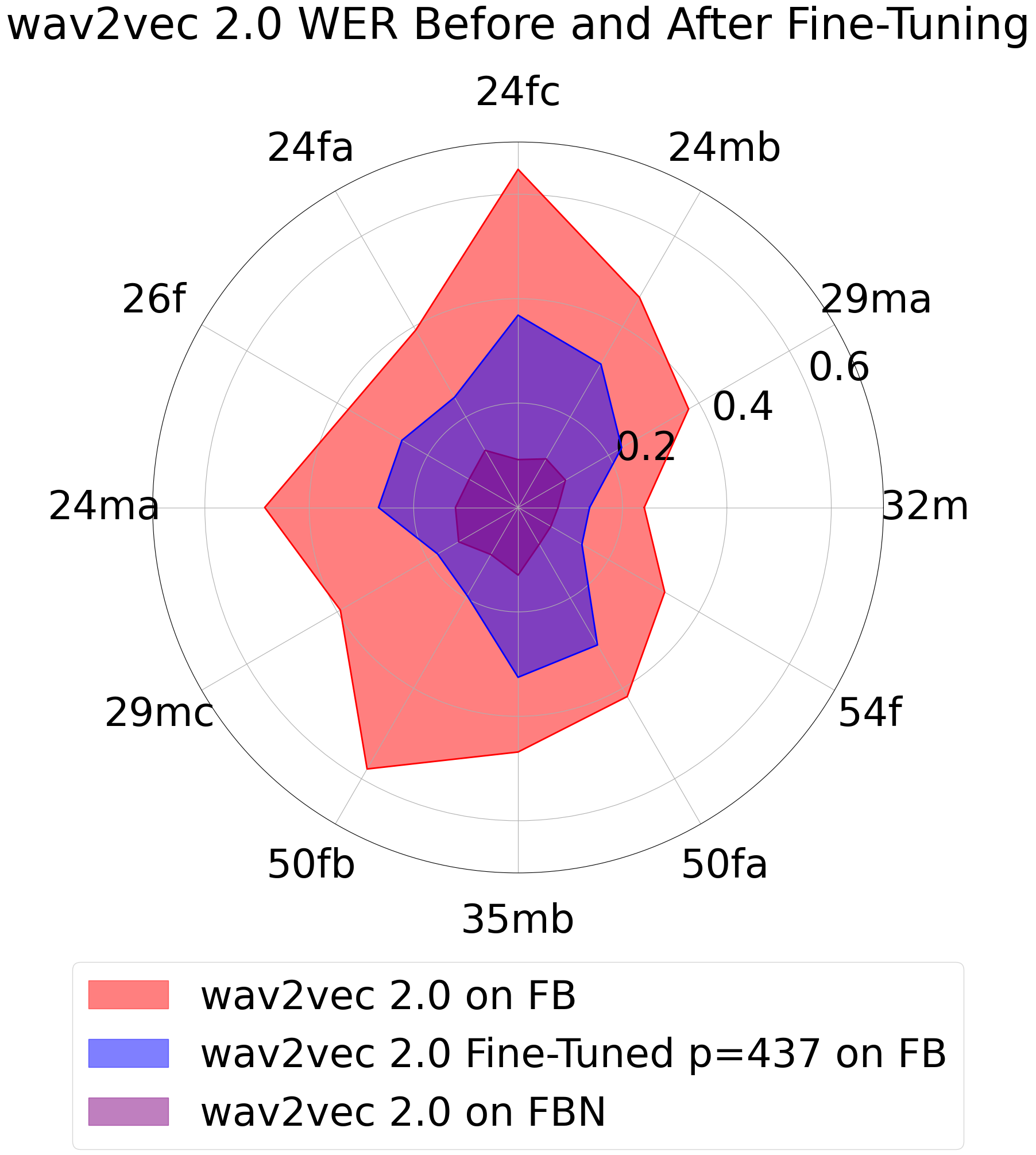}
\caption{Plot illustrating WER per speaker on FluencyBank (FB) before and after model fine-tuning with data augmentation, as well as model accuracy on non-disfluent speech, FluencyBank-N (FBN). Speakers are represented by their age and gender, as identified in the FluencyBank dataset. Thus, ``26f'' represents a 26-year old female.}
\label{fig:errorarea}
\vspace*{-.2in}
\end{figure}

The results, detailed in Table \ref{tbl:metrics}, indicate a positive trend: incorporating augmented disfluent speech samples during training leads to further improvements in WER and $F_{BERT}$. However, we note a minor increase in WER for FluencyBank-N, suggesting a potential overcorrection towards stuttered speech characteristics. Despite this, as the training incorporates a larger pool of samples, the WER begins to stabilize and shows a gradual decline, underlining the benefits of extensive data augmentation.

Our analysis across different ASR configurations and FluencyBank video types---interviews and reading passages---revealed variable WER performance. Interviews, involving personal experience discussions, showed less ground truth word overlap than reading passages, yet occasionally had content overlap due to discussions on life with stuttering. Detailed in Table \ref{tbl:videotypemetrics}, this analysis mirrors our overall results (Table \ref{tbl:metrics}), with reading passages experiencing greater WER improvements than interviews, though both video types saw notable enhancements.

We also analyzed the WER across different speakers and disfluency types. The speaker-specific WERs for three distinct model configurations are shown in Figure~\ref{fig:errorarea}, and the WER distributions related to specific disfluencies for four configurations are presented in Figure~\ref{fig:dist_plot_disfluencies}. Our focus was on repetitions and interjections, identified through analysis of CHAT codes in FluencyBank transcripts. We found that 31\% of utterances in the FluencyBank dataset contained at least one word and/or phrase repetition, and 59\% included interjections, though other disfluency types were also noted. Our primary analysis remained focused on these disfluencies, which were also the key targets in our data augmentation process. Post fine-tuning and data augmentation, we observed a reduction in WER across all speakers and various age and gender demographics. Additionally, there was a decrease in disfluency-specific WER for each model configuration, demonstrating the efficacy of our approach.

\section{Conclusion}
In this paper, we introduced a new approach to improving ASR accuracy for individuals who stutter, overcoming previous limitations in dataset size and diversity with a novel data augmentation strategy that infuses new training samples with synthetic disfluencies. Our findings reveal that even minimal fine-tuning with disfluent speech significantly enhances ASR performance, as evidenced by reduced WER, with further improvements stemming from our data augmentation techniques.

Looking ahead, we plan to broaden our research to include a wider array of disfluency types, aiming for a more comprehensive representation of stuttered speech. We also seek to fine-tune our methodology to better align ASR-generated transcripts with the preferences of people who stutter, acknowledging that some may wish to see their disfluencies reflected in transcriptions to accurately convey their speech patterns. This aspiration underscores the importance of understanding the diverse needs and preferences of those who stutter. Ultimately, our research contributes to making ASR technology more accessible and adaptable, laying the groundwork for future innovations that accommodate a range of speech variations.

\section{Acknowledgements}
This material is based upon work supported by the U.S. National Science Foundation under Grant Nos. 2235916 and 2345086.

\bibliographystyle{IEEEtran}
\bibliography{mybib}

\begin{thebibliography}{10}
\providecommand{\url}[1]{#1}
\csname url@samestyle\endcsname
\providecommand{\newblock}{\relax}
\providecommand{\bibinfo}[2]{#2}
\providecommand{\BIBentrySTDinterwordspacing}{\spaceskip=0pt\relax}
\providecommand{\BIBentryALTinterwordstretchfactor}{4}
\providecommand{\BIBentryALTinterwordspacing}{\spaceskip=\fontdimen2\font plus
\BIBentryALTinterwordstretchfactor\fontdimen3\font minus
  \fontdimen4\font\relax}
\providecommand{\BIBforeignlanguage}[2]{{%
\expandafter\ifx\csname l@#1\endcsname\relax
\typeout{** WARNING: IEEEtran.bst: No hyphenation pattern has been}%
\typeout{** loaded for the language `#1'. Using the pattern for}%
\typeout{** the default language instead.}%
\else
\language=\csname l@#1\endcsname
\fi
#2}}
\providecommand{\BIBdecl}{\relax}
\BIBdecl

\bibitem{tichenor2019stuttering}
S.~E. Tichenor and J.~S. Yaruss, ``Stuttering as defined by adults who
  stutter,'' \emph{Journal of Speech, Language, and Hearing Research}, vol.~62,
  no.~12, pp. 4356--4369, 2019.

\bibitem{wu2023world}
S.~Wu, ````{The World is Designed for Fluent People}'': Benefits and challenges
  of videoconferencing technologies for people who stutter,'' in
  \emph{Proceedings of the 2023 CHI Conference on Human Factors in Computing
  Systems}, 2023, pp. 1--17.

\bibitem{johnson1959onset}
W.~Johnson, \emph{The onset of stuttering: Research findings and
  implications}.\hskip 1em plus 0.5em minus 0.4em\relax U of Minnesota Press,
  1959.

\bibitem{yairi2013epidemiology}
E.~Yairi and N.~Ambrose, ``Epidemiology of stuttering: 21st century advances,''
  \emph{Journal of fluency disorders}, vol.~38, no.~2, pp. 66--87, 2013.

\bibitem{tichenor2021variability}
S.~E. Tichenor and J.~S. Yaruss, ``Variability of stuttering: Behavior and
  impact,'' \emph{American Journal of Speech-Language Pathology}, vol.~30,
  no.~1, pp. 75--88, 2021.

\bibitem{tobin2022personalized}
J.~Tobin and K.~Tomanek, ``Personalized automatic speech recognition trained on
  small disordered speech datasets,'' in \emph{ICASSP 2022-2022 IEEE
  International Conference on Acoustics, Speech and Signal Processing
  (ICASSP)}.\hskip 1em plus 0.5em minus 0.4em\relax IEEE, 2022, pp. 6637--6641.

\bibitem{lea2023user}
C.~Lea, Z.~Huang, J.~Narain, L.~Tooley, D.~Yee, D.~T. Tran, P.~Georgiou, J.~P.
  Bigham, and L.~Findlater, ``From user perceptions to technical improvement:
  Enabling people who stutter to better use speech recognition,'' in
  \emph{Proceedings of the 2023 CHI Conference on Human Factors in Computing
  Systems}, 2023, pp. 1--16.

\bibitem{mujtaba2019ethical}
D.~F. Mujtaba and N.~R. Mahapatra, ``Ethical considerations in {AI}-based
  recruitment,'' in \emph{2019 IEEE International Symposium on Technology and
  Society (ISTAS)}.\hskip 1em plus 0.5em minus 0.4em\relax IEEE, 2019, pp.
  1--7.

\bibitem{gerlach2018stuttering}
H.~Gerlach, E.~Totty, A.~Subramanian, and P.~Zebrowski, ``Stuttering and labor
  market outcomes in the united states,'' \emph{Journal of Speech, Language,
  and Hearing Research}, vol.~61, no.~7, pp. 1649--1663, 2018.

\bibitem{plexico2019influence}
L.~W. Plexico, M.-B. Hamilton, H.~Hawkins, and S.~Erath, ``The influence of
  workplace discrimination and vigilance on job satisfaction with people who
  stutter,'' \emph{Journal of Fluency Disorders}, vol.~62, p. 105725, 2019.

\bibitem{techexplore}
K.~Wheeler, ``For people who stutter, the convenience of voice assistant
  technology remains out of reach,''
  \url{https://techxplore.com/news/2020-01-people-stutter-convenience\\-voice-technology.html
  }.

\bibitem{sheikh2022machine}
S.~A. Sheikh, M.~Sahidullah, F.~Hirsch, and S.~Ouni, ``Machine learning for
  stuttering identification: Review, challenges and future directions,''
  \emph{Neurocomputing}, 2022.

\bibitem{lea2021sep}
C.~Lea, V.~Mitra, A.~Joshi, S.~Kajarekar, and J.~P. Bigham, ``{SEP}-28k: A
  dataset for stuttering event detection from podcasts with people who
  stutter,'' in \emph{ICASSP 2021-2021 IEEE International Conference on
  Acoustics, Speech and Signal Processing (ICASSP)}.\hskip 1em plus 0.5em minus
  0.4em\relax IEEE, 2021, pp. 6798--6802.

\bibitem{kourkounakis2020fluentnet}
T.~Kourkounakis, A.~Hajavi, and A.~Etemad, ``{FluentNet}: end-to-end detection
  of speech disfluency with deep learning,'' \emph{arXiv preprint
  arXiv:2009.11394}, 2020.

\bibitem{mitra2021analysis}
V.~Mitra, Z.~Huang, C.~Lea, L.~Tooley, S.~Wu, D.~Botten, A.~Palekar,
  S.~Thelapurath, P.~Georgiou, S.~Kajarekar \emph{et~al.}, ``Analysis and
  tuning of a voice assistant system for dysfluent speech,'' \emph{arXiv
  preprint arXiv:2106.11759}, 2021.

\bibitem{moore2020uncommonvoice}
M.~Moore, P.~Papreja, M.~Saxon, V.~Berisha, and S.~Panchanathan,
  ``{UncommonVoice}: A crowdsourced dataset of dysphonic speech.'' in
  \emph{Interspeech}, 2020, pp. 2532--2536.

\bibitem{hermannfew}
E.~Hermann and M.~M. Doss, ``Few-shot dysarthric speech recognition with
  text-to-speech data augmentation.''

\bibitem{baskar2022speaker}
M.~K. Baskar, T.~Herzig, D.~Nguyen, M.~Diez, T.~Polzehl, L.~Burget, and
  J.~{\v{C}}ernock{\`y}, ``Speaker adaptation for wav2vec2 based dysarthric
  {ASR},'' \emph{arXiv preprint arXiv:2204.00770}, 2022.

\bibitem{macwhinney2024talkbank}
B.~Macwhinney and D.~Fromm, ``{TalkBank} methods for studying spoken
  discourse,'' in \emph{Spoken Discourse Impairments in the Neurogenic
  Populations: A State-of-the-Art, Contemporary Approach}.\hskip 1em plus 0.5em
  minus 0.4em\relax Springer, 2024, pp. 97--109.

\bibitem{ratner2018fluency}
N.~B. Ratner and B.~MacWhinney, ``Fluency {Bank}: A new resource for fluency
  research and practice,'' \emph{Journal of fluency disorders}, vol.~56, pp.
  69--80, 2018.

\bibitem{radford2023robust}
A.~Radford, J.~W. Kim, T.~Xu, G.~Brockman, C.~McLeavey, and I.~Sutskever,
  ``Robust speech recognition via large-scale weak supervision,'' in
  \emph{International Conference on Machine Learning}.\hskip 1em plus 0.5em
  minus 0.4em\relax PMLR, 2023, pp. 28\,492--28\,518.

\bibitem{wav2vecweights}
Facebook, ``wav2vec2-base-960h,''
  \url{https://huggingface.co/facebook/wav2vec2-base-960h}.

\bibitem{panayotov2015librispeech}
V.~Panayotov, G.~Chen, D.~Povey, and S.~Khudanpur, ``Libri{S}peech: an {ASR}
  corpus based on public domain audio books,'' in \emph{Acoustics, Speech and
  Signal Processing (ICASSP), 2015 IEEE International Conference on}.\hskip 1em
  plus 0.5em minus 0.4em\relax IEEE, 2015, pp. 5206--5210.

\bibitem{macwhinney2017tools}
B.~MacWhinney, ``Tools for analyzing talk part 1: The {CHAT} transcription
  format,'' \emph{Carnegie.[Google Scholar]}, vol.~16, 2017.

\bibitem{ao-etal-2022-speecht5}
J.~Ao, R.~Wang, L.~Zhou, C.~Wang, S.~Ren, Y.~Wu, S.~Liu, T.~Ko, Q.~Li,
  Y.~Zhang, Z.~Wei, Y.~Qian, J.~Li, and F.~Wei, ``{S}peech{T}5: Unified-modal
  encoder-decoder pre-training for spoken language processing,'' in
  \emph{Proceedings of the 60th Annual Meeting of the Association for
  Computational Linguistics (Volume 1: Long Papers)}, May 2022, pp. 5723--5738.

\bibitem{cmuvectors}
M.~Hollemans, ``Mattijs/cmu-arctic-xvectors,''
  \url{https://huggingface.co/datasets/Matthijs/cmu-arctic-xvectors}.

\bibitem{openai}
``{OpenAI Text-to-Speech Service},''
  \url{https://platform.openai.com/docs/models/tts}, accessed: February 2024.

\bibitem{tobin2022assessing}
J.~Tobin, Q.~Li, S.~Venugopalan, K.~Seaver, R.~Cave, and K.~Tomanek,
  ``Assessing {ASR} model quality on disordered speech using {BERTScore},''
  \emph{arXiv preprint arXiv:2209.10591}, 2022.

\bibitem{zhang2019bertscore}
T.~Zhang, V.~Kishore, F.~Wu, K.~Q. Weinberger, and Y.~Artzi, ``{BERTScore}:
  Evaluating text generation with {BERT},'' \emph{arXiv preprint
  arXiv:1904.09675}, 2019.

\bibitem{huggingface_deberta_large_mnli}
\BIBentryALTinterwordspacing
P.~He, X.~Liu, J.~Gao, and W.~Chen, ``De{BERT}a: Decoding-enhanced {BERT} with
  disentangled attention,'' in \emph{International Conference on Learning
  Representations}, 2021. [Online]. Available:
  \url{https://openreview.net/forum?id=XPZIaotutsD}
\BIBentrySTDinterwordspacing

\bibitem{baevski2020wav2vec}
A.~Baevski, Y.~Zhou, A.~Mohamed, and M.~Auli, ``wav2vec 2.0: A framework for
  self-supervised learning of speech representations,'' \emph{Advances in
  neural information processing systems}, vol.~33, pp. 12\,449--12\,460, 2020.

\bibitem{normalizer}
HuggingFace, ``Normalization and pre-tokenization,''
  \url{https://huggingface.co/learn/nlp-course/chapter6/4}.

\end{thebibliography}

\end{document}